\documentclass[letterpaper]{article} % DO NOT CHANGE THIS
\usepackage{aaai24}  % DO NOT CHANGE THIS
\usepackage{times}  % DO NOT CHANGE THIS
\usepackage{helvet}  % DO NOT CHANGE THIS
\usepackage{courier}  % DO NOT CHANGE THIS
\usepackage[hyphens]{url}  % DO NOT CHANGE THIS
\usepackage{graphicx} % DO NOT CHANGE THIS
\usepackage{natbib}  % DO NOT CHANGE THIS AND DO NOT ADD ANY OPTIONS TO IT
\usepackage{caption} % DO NOT CHANGE THIS AND DO NOT ADD ANY OPTIONS TO IT
\usepackage{algorithm}
\usepackage{algorithmic}

\usepackage[table]{xcolor}
\usepackage{subcaption}
\usepackage{graphicx}
\usepackage[utf8]{inputenc} % allow utf-8 input
\usepackage[T1]{fontenc}    % use 8-bit T1 fonts
\usepackage{booktabs}       % professional-quality tables
\usepackage{amsfonts}       % blackboard math symbols
\usepackage{nicefrac}       % compact symbols for 1/2, etc.
\usepackage{microtype}      % microtypography
\usepackage{xcolor}         % colors
\usepackage[english]{babel}
\usepackage{amsthm}
\usepackage{amsmath}
\theoremstyle{definition}
\newtheorem{definition}{Threat model}
\usepackage{multirow}
\usepackage{url}

\usepackage{newfloat}
\usepackage{listings}
\DeclareCaptionStyle{ruled}{labelfont=normalfont,labelsep=colon,strut=off} % DO NOT CHANGE THIS
\floatstyle{ruled}
\newfloat{listing}{tb}{lst}{}
\floatname{listing}{Listing}
\title{Towards Deep Learning Models Resistant to Transfer-based Adversarial Attacks via Data-centric Robust Learning}
\author{
    Yulong Yang \textsuperscript{\rm 1},
    Chenhao Lin \thanks{Corresponding author.} \textsuperscript{\rm 1}, 
    Xiang Ji \textsuperscript{\rm 1},
    Qiwei Tian \textsuperscript{\rm 1},
    Qian Li \textsuperscript{\rm 1},
    Hongshan Yang \textsuperscript{\rm 2},
    Zhibo Wang \textsuperscript{\rm 2},
    Chao Shen \textsuperscript{\rm 1}, \\   
}
\affiliations{
    \textsuperscript{\rm 1}School of Cyber Science and Engineering, Xi'an Jiaotong University, P.R. China \\
    \textsuperscript{\rm 2}School of Cyber Science and Technology, Zhejiang University, P.R. China \\
    \{xjtu2018yyl0808, xiangji, michaeltqw\}@stu.xjtu.edu.cn, 
    \{linchenhao, qianlix, chaoshen\}@xjtu.edu.cn, \{yanghs, zhibowang\}@zju.edu.cn
}

\usepackage{bibentry}
\begin{document}

\maketitle

\begin{abstract}
Transfer-based adversarial attacks raise a severe threat to real-world deep learning systems since they do not require access to target models. Adversarial training (AT), which is recognized as the strongest defense against white-box attacks, has also guaranteed high robustness to (black-box) transfer-based attacks.
However, AT suffers from heavy computational overhead since it optimizes the adversarial examples during the whole training process.
In this paper, we demonstrate that such heavy optimization is not necessary for AT against transfer-based attacks. 
Instead, a one-shot adversarial augmentation prior to training is sufficient, and we name this new defense paradigm Data-centric Robust Learning (DRL).
Our experimental results show that DRL outperforms widely-used AT techniques (e.g., PGD-AT, TRADES, EAT, and FAT) in terms of black-box robustness and even surpasses the top-1 defense on RobustBench when combined with diverse data augmentations and loss regularizations.
We also identify other benefits of DRL, for instance, the model generalization capability and robust fairness.
\end{abstract}

\section{Introduction}
Recent years have witnessed a rapid development of Deep Neural Networks (DNNs) 
\cite{lecun2015deep} on tasks like image processing, language understanding, etc. However, DNNs are found to be vulnerable to adversarial examples 
\cite{szegedy2013intriguing, biggio2013evasion}, which are crafted by adding human-imperceptible perturbations and can mislead the prediction of DNNs. The existence of adversarial examples brings security issues to DNNs and hinders their wider application.

The adversary can craft adversarial examples to attack the DNNs under different threat models. Among the various threat models, two prominent ones are extensively discussed: white-box attacks \cite{madry2017towards} and black-box attacks \cite{papernot2017practical}. White-box attacks assume the adversary has access to all the information related to the target DNN, including the training data, the network architecture, the gradient of the model, etc. These attacks are used to test the robustness of DNNs under extreme cases and are usually not applicable in the real world 
\cite{papernot2017practical}. In contrast, black-box attacks assume the adversary only has access to partial information about the target DNNs, such as the training dataset (referred to as transfer-based attacks) or the model outputs (referred to as query-based attacks).

\begin{figure}[t]
    \centering
    \includegraphics[width=.48\textwidth]{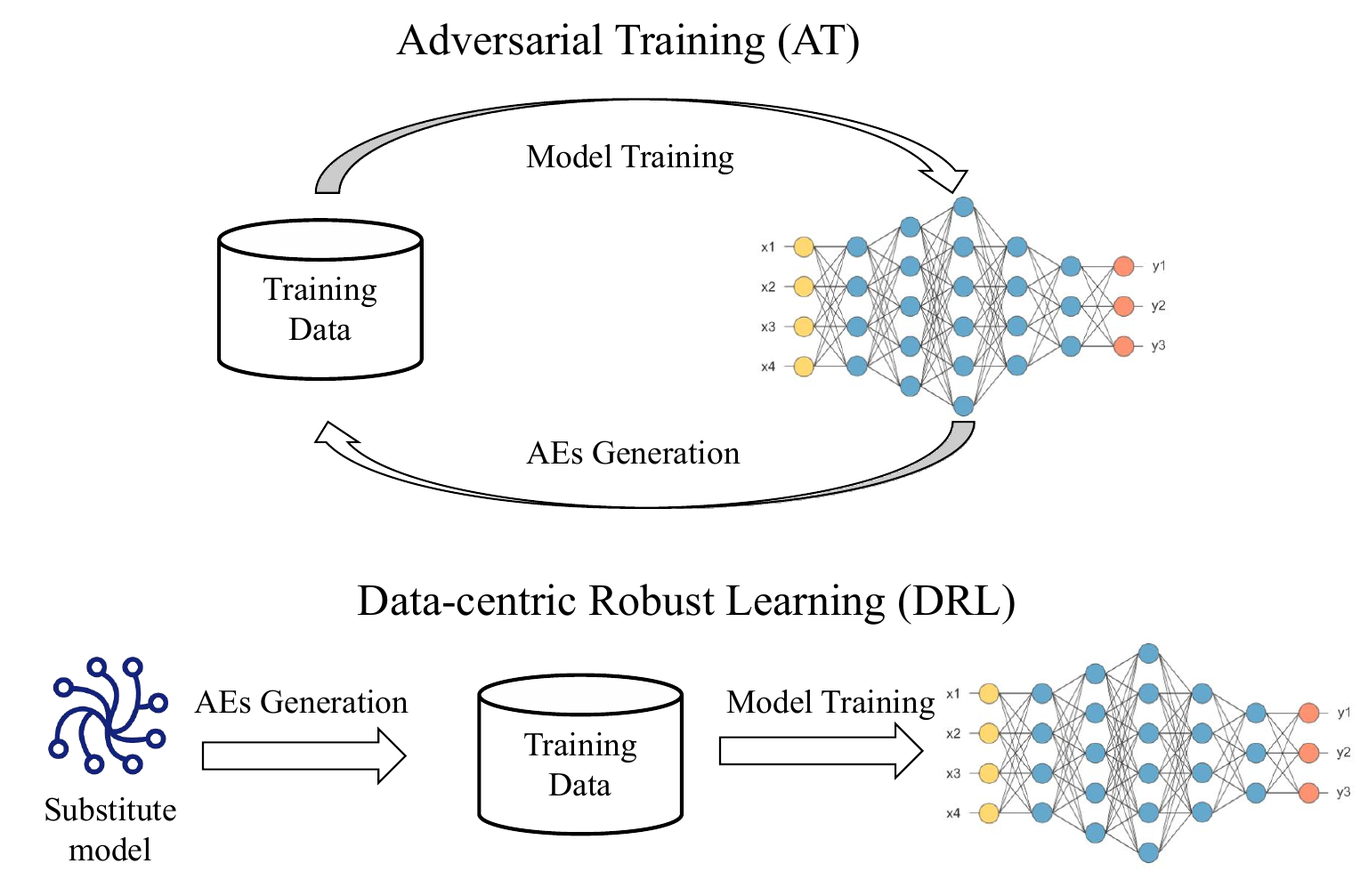}
    \caption{The training paradigm of Adversarial Training (AT) vs. our Data-centric Robust Learning (DRL).}
    \label{fig:AT-vs-DRL}
\end{figure}

State-of-the-art (SOTA) black-box attacks have demonstrated remarkable attack performance. For instance, Zhu et al. \cite{zhu2022toward} and Li et al. \cite{li2023making} achieved transfer-based attack success rates of over 90\% across commonly used Convolutional Neural Networks (CNNs), and Mou et al. ~ \cite{Mou2022AFS} achieved 100\% success rates in query-based attacks against commercial computer vision APIs within only 20 queries. Therefore, defense methods are urgently needed to mitigate these SOTA black-box attacks.

Several recent works have developed defense methods against black-box attacks with either pre-processing \cite{qin2021random}, post-processing \cite{chen2022adversarial}, or query statistics \cite{li2022blacklight}. However, a common limitation of these defenses is that they cannot effectively defend against transfer-based attacks. In realistic black-box settings, the adversary can apply either transfer-based or query-based attacks to the target system. Defenses against transfer-based attacks are thus of significance in achieving comprehensive protection under the black-box settings. Unfortunately, recent studies have paid little attention to this aspect of defense, to our best knowledge.

Adversarial training \cite{madry2017towards, zhang2019theoretically} is a general defense under the white-box setting and has been shown to be effective in defending against transfer-based black-box attacks. However, this defense strategy has several limitations, hindering its deployment in the realistic black-box setting. First, adversarial training is computationally expensive due to the large amount of white-box adversarial examples generated during training. Second, adversarial training harms the model's generalization ability. Therefore, this paper aims to find an alternative and more affordable solution to mitigate transfer-based black-box attacks. 

This paper finds that transfer-based black-box attacks can be easily mitigated with a one-shot training paradigm, which generates the augmented adversarial examples only before the training stage and keeps the whole dataset unchanged during the training period, as illustrated in Fig.\ref{fig:AT-vs-DRL}. On the basis of the above insights, this paper proposes Data-centric Robust Learning (DRL) to generate a high-quality dataset to improve the black-box robustness of the DNNs. DRL generates the augmented dataset in the novel one-shot paradigm and adopts a data selection mechanism based on the output confidence of the training example to accelerate the convergence of the training. Experimental results on three datasets, including CIFAR-10, CIFAR-100, and TinyImageNet, demonstrate that DRL achieves higher robustness against transfer-based adversarial attacks than adversarial training and is much more computationally efficient. Additionally, DRL exhibits higher clean accuracy and Out-Of-Distribution (OOD) robustness and is much class-wise fairer than adversarial training. Moreover, DRL can be further combined with recently proposed data augmentation methods and black-box defenses to achieve both SOTA black-box adversarial robustness and OOD robustness. 

In summary, our contributions are as follows.
\begin{itemize}
% comment: 强调一下DRL新的训练范式
    \item We find that transfer-based black-box attacks can be easily mitigated with a one-shot training paradigm, which is a more efficient and affordable solution compared to adversarial training methods.
    \item We propose a one-shot dataset augmentation paradigm DRL, which consists of clean examples augmentation, adversarial examples augmentation, and data selection to close the gap in mitigating transfer-based black-box attacks. 
    \item DRL can be combined with other black-box defenses and common techniques, such as data augmentations, to achieve comprehensive protection under the realistic black-box setting. 
\end{itemize}

\section{Related Work}
\textbf{Black-box attacks and defenses.} Black-box adversarial attacks are usually divided into transfer-based attacks 
 \cite{papernot2017practical,zhu2022toward,li2023making} and query-based attacks \cite{chen2017zoo,Mou2022AFS}. The transfer-based attacks assume the adversary has access to the training dataset and can train a local substitute model to generate adversarial examples. The query-based attacks assume the adversary can query the target DNN and get the output score or output label, which can be used to approximate the attack gradients. Many defenses against query-based attacks have been proposed recently. For instance, Blacklight \cite{li2022blacklight} detected malicious attack queries by analyzing the query statistics of the system. \citet{qin2021random} and \citet{chen2022adversarial} proposed to disturb the gradient approximation by adding noise on either input examples or output scores. While these methods are effective defenses against query-based black-box attacks, they are not robust against transfer-based attacks. 

\textbf{Data-centric machine learning.} Data-centric machine learning is advocated by Andrew Ng \cite{zha2023data} et al., which focuses on creating high-quality datasets rather than designing more sophisticated model architectures to improve the performance of DNNs. Several methods proposed in the research community can be regarded as data-centric approaches, such as data augmentation and adversarial training \cite{zha2023data}. Recently, \citet{faghri2023reinforce} proposed distilling the dataset with large pre-trained teacher models to enhance the clean accuracy, OOD robustness, and transfer learning performance. However, there is still a lack of studies focusing on improving the robustness of DNNs against transfer-based black-box adversarial attacks, especially from the perspective of enhancing the dataset. 

\textbf{Adversarial training.} Adversarial training \cite{madry2017towards, zhang2019theoretically,tramer2017ensemble,rebuffi2021fixing,wang2023better} is widely considered the SOTA defense against white-box adversarial attacks. At the training stage, adversarial training iteratively generates adversarial examples and trains the model to be robust against them. Although adversarial training achieves high white-box robustness and can also enhance the black-box robustness of DNNs, the heavy computational overhead impedes its deployment in real settings. This paper provides a more affordable solution in defending realistic transfer-based black-box attacks with the one-shot training paradigm.

\textbf{Data augmentation.} Data augmentation \cite{hendrycks2019augmix, wang2021augmax} is used to improve the generalization of DNNs on unseen test data and corrupted data, such as those containing noises, blurs, snow, frost, fog, JPEG compression corruptions, etc. Various data augmentation techniques are also adopted to enhance the performance of Adversarial Training (AT). For example, SOTA AT methods commonly use generative models to expand the dataset \cite{rebuffi2021fixing,wang2023better}.

\textbf{Adversarial defense based on detection or purification.} Adversarial detection methods use detectors to judge whether an input example is adversarial or not. The detector can be trained with either supervised \cite{smith2018understanding,monteiro2019generalizable,zuo2021exploiting} or unsupervised \cite{sheikholeslami2020minimum,aldahdooh2023revisiting} manner. Adversarial purification methods defend the adversarial attacks by purifying the perturbed examples before classification. Recently proposed adversarial purification methods \cite{xu2022general,nie2022diffusion,nayak2022data} purify the adversarial examples with generative models. For example, \citet{nie2022diffusion} purify adversarial examples with diffusion models \cite{song2020score}. Please note that adversarial detection and purification methods are plug-n-play and can be combined with our DRL method to achieve higher robustness.

\section{Methodology}
\label{sec:methodology}
\subsection{Threat Models}
This paper mainly focuses on the realistic transfer-based black-box threat model, where the adversary has limited knowledge of the target DNN architecture and defense methods and can only use the clean training dataset to train substitute models. We will also define the adaptive attack threat model in Sec.\ref{sec:adaptive-evaluation} to stress test the defense effectiveness when the black-box settings are broken and the adversary has the extra knowledge.

\begin{definition}[Realistic Attack]
The adversary only knows the distribution of the original dataset $\mathcal{D}$ but does not know the augmented dataset $\mathcal{D}'$, the target DNN architecture $M$, and the training loss $\mathcal{L}$.
\end{definition}

\subsection{DRL: Data-centric Robust Learning for Defending Transfer-based Attacks}

\textbf{Problem formulation.} Given an original training distribution $D$, a fixed training algorithm $train()$ and a fixed architecture $H$, the aim of DRL is to generate an augmented distribution $D'$, such that the trained model $h\in H$ has high accuracy on clean distribution $D$ and unseen adversarial distribution $D^{A}$, that is,
\begin{equation}
    \mathop{\arg\min}\limits_{h} \sum_{i=1}^{N}\mathbb{I}(h(x_i)\neq y_i),
\end{equation}
where $\{x_i, y_i\}$ are sampled from distribution $D \cup D^{A}$ and $h=train(H, D')$; $\mathbb{I}$ is the indicator function. This paper considers the adversarial examples constrained with $l_{\infty}$-norm, that is, $||x_{adv}-x||_{\infty}\leq \epsilon$. The proposed DRL approach consists of two parts: adversarial data augmentation and data selection. Adversarial data augmentation enhances the model robustness by expanding the training dataset, while the data selection method is necessary to prevent the model from overfitting and is beneficial to the robustness improvement, as verified in Sec.\ref{sec:ablation}.

\textbf{Adversarial data augmentation.} In the adversarial data generation stage, we generate transferable adversarial examples to augment the dataset. The substitute models are normally DNNs with the same architecture as the DNNs to be trained. To make a fair comparison with AT methods, DRL only adopts the PGD algorithm to generate adversarial examples, which can be formalized as,
\begin{equation}
    x^{t+1} = \Pi_{\epsilon} (x^t +  \alpha \cdot sign(\nabla_x \mathcal{L}(f_\theta, x^t, y))),
\end{equation}
where $f_{\theta}$ is the substitute model, $\Pi_{\epsilon}$ is the projection operator, and $\mathcal{L}$ is a classification loss, e.g., cross-entropy loss, $x^{t}$ is the adversarial examples generated at the $t$-step and $\alpha$ is the step size.

\textbf{Data selection.} We design a data selection method to prevent the model from overfitting and enhance the DRL effectiveness. Given the DRL dataset consisting of N examples, we only select M (M<N) examples for each training epoch. For example, on the CIFAR-10 dataset, we select only 50,000 images per epoch. Specifically, we adopt the output confidence gap metric to select the training example, that is,
\begin{equation}
    M_{conf} = \frac{1}{2}[(F(x)_{y} - \max_{i \neq y} F(x)_{i}) + (F(x')_{y} - \max_{i \neq y} F(x')_{i})],
\end{equation}
where $F$ is the model with the softmax output layer; $F(x)_{i}$ denotes the softmax value corresponding to the $i$-th class; $x, x', y$ is the clean example, its corresponding adversarial examples, and true label. Training data with lower $M_{conf}$ value has higher priority to be used. DRL initializes the $M_{conf}$ to zero. After each training batch, the $M_{conf}$ of the examples in the current batch will be updated. Each epoch selects the top $M$ training examples in increasing order of $M_{conf}$. Intuitively, this data selection criterion encourages the data with lower confidence to be used in the next epoch, thus prevent model from overfitting to ``easy'' training samples.

\textbf{Loss function.} To highlight the effectiveness of the proposed data-centric one-shot training paradigm, DRL directly adopts Cross-Entropy (CE) loss to optimize. Note that the design of loss function belongs to model-centric technique, which can be combined with DRL to promote the robustness.

\subsection{XDRL: Combining DRL with Other Common Techniques}

In Sec.\ref{sec:comparing-with-sota-at}, we will show that DRL can easily outperform SOTA AT methods in defending transfer-based black-box attacks by simply combining with the following common techniques (denoted as XDRL in the following).

\textbf{Synthetic data augmentation.} Using generative models to expand the dataset has been proven to be effective in enhancing the robustness of adversarial training~ \cite{rebuffi2021fixing,wang2023better}. This technique can also be applied to the XDRL. Specifically, we follow \cite{wang2023better} to generate $M$ clean data with the class-conditional Elucidating Diffusion Model (EDM). We keep each class evenly distributed in the generated synthetic dataset.

\textbf{Attack augmentation} The diversity of adversarial examples can be augmented by applying various attack algorithms. Specifically, XDRL adopts four adversarial attacks, including FGSM \cite{goodfellow2014explaining}, PGD \cite{madry2017towards}, C\&W \cite{carlini2017towards}, and MIM \cite{dong2018boosting}. Specifically, FGSM generates adversarial example $x^{adv}$ with a single-step optimization,
\begin{equation}
    x^{adv} = x + \epsilon \cdot sign(\nabla_x \mathcal{L}(f_\theta, x, y)),
\end{equation}
where $f_{\theta}$ is the substitute model with parameter $\theta$, and $\mathcal{L}$ is a classification loss, e.g., cross-entropy loss. C\&W iteratively generates adversarial examples by minimizing the following optimization objective,
\begin{equation}
    \mathcal{L} = ReLU(\max_{i \neq c}(f(x)_i) - f(x)_c + \kappa),
\end{equation}
where $f(x)_i$ denotes the logit value of the $i$-th class, $c$ is the wrong class with the largest logit value, and $\kappa>0$ controls the attack hardness. MIM stabilizes the attack by adding a momentum term,
\begin{gather}
    x^{t+1} = x^t + \alpha \cdot sign(g_{t+1}), \\
    g^{t+1} = g^t + \frac{\nabla_x \mathcal{L}(f_\theta, x_{t}, y)}{||\nabla_x \mathcal{L}(f_\theta, x_{t}, y)||_1}.
\end{gather}

\textbf{Alignment regularization loss function.} After augmenting the dataset, we can train the robust models using the following training objective, which combines CE loss with an alignment regularization term,
\begin{equation}
\label{eqn:naive-DRL}
    \min_{f_\theta} \ \mathbb{E}_{(x, x^{adv}, y)\sim \mathcal{D'}} \ \mathcal{L}_{ce}(f_\theta, x, y) + \lambda \mathcal{L}_c(f_{\theta}, x, x^{adv}),
\end{equation}
where $\mathcal{L}_{ce}$ is the cross-entropy loss, $\mathcal{L}_c$ is the consistency loss, and $\lambda$ is the regularization hyper-parameter. We have tried three commonly used regularization terms, including the $l_{1}$-norm, squared $l_{2}$-norm, and KL-divergence. We find that $l_{1}$-norm is the most effective one (See Sec. \ref{sec:ablation}).

\textbf{Corrupted data augmentation.} When combined with naive corrupted Data Augmentation (DA) and SOTA methods like AugMix \cite{hendrycks2019augmix} and AugMax \cite{wang2021augmax}, DRL can further enhance the robustness. Specifically, when combined with corrupted data augmentation methods, we simply replace the clean data $x$ term with the DRL data $x'$ and keep the rest of the training procedure consistent with the original data augmentation method. For instance, combining DRL with DA can be formalized as,
\begin{equation}
    \min_{f_\theta} \ \mathbb{E}_{(x', y)\sim \mathcal{D'}} \  \mathcal{L}_{ce}(f_\theta,T(x'), y),
\end{equation}
where $T$ is the data transformation function used to augment data, $x'$ is the data sampled from the DRL dataset $\mathcal{D}'$. Combining DRL with AugMix can be formalized as,
\begin{equation}
    \min_{f_\theta} \ \mathbb{E}_{(x', y)\sim \mathcal{D'}} \ \mathcal{L}_{ce}(f_\theta,x',y) + \lambda \mathcal{L}_c(f_\theta,x,T(x')),
\end{equation}
where $\mathcal{L}_c$ is the consistency loss and $\lambda$ is the regularization hyper-parameter. Combining DRL with AugMax can be formalized as,
\begin{equation}
\begin{aligned}
    \min_{f_\theta} \ \mathbb{E}_{(x', y)\sim \mathcal{D'}} \  \frac{1}{2}\left[\mathcal{L}_{ce}(f_\theta, T(x'), y) + \mathcal{L}_{ce}(f_\theta, x', y)\right] \\
    + \lambda \mathcal{L}_c(x', T(x')). 
\end{aligned}
\end{equation}

\section{Experimental evaluation}
\label{sec:evaluation}

\subsection{Experimental setup}

\textbf{Dataset.} We select the popular image datasets CIFAR-10, CIFAR-100, and TinyImageNet for evaluation. We show the results on CIFAR-10 in this section, while results on other datasets can be found in the \textit{Supplementary File}.

\textbf{Model architecture.} We adopt two popular architectures WideResNet-28-10 
 \cite{zagoruyko2016wide} and PreActResNet-18 \cite{he2016deep} for a convenient comparison with other adversarial training (AT) methods. We also adopt the Res18-DuBIN model for a fair comparison with AugMax \cite{wang2021augmax}.

\textbf{Robustness evaluation.} We generate transferable adversarial examples on the test set to evaluate the black-box adversarial robustness of DRL. The substitute model used is normally trained ResNet-34. We set the perturbation budget $\epsilon = 8/255$ under the $l_\infty$-norm constraint. We adopt both in-box attacks (PGD) and out-of-box attacks (C\&W, QAA \cite{yang2023quantization}, and Ens Attack). Ens Attack denotes attacking ensembled substitute models to enhance the attack transferability. Please note that we do not apply other transfer-based attacks such as DIM \cite{xie2019improving}, TIM \cite{dong2019evading}, FIA \cite{wang2021feature}, and RPA \cite{zhang2022enhancing} in the experiments because QAA has demonstrated superior performance compared to these methods. More details on these transfer attacks can be found in the \textit{Supplementary File}. We use Robust Accuracy (RA), i.e., the accuracy of the transferable adversarial examples, as the robustness evaluation metric.

\textbf{Robust fairness evaluation.} We adopt the class-wise accuracy standard deviation (the standard deviation of the accuracy per class) to evaluate the robust fairness of the model, and the lower is better.

\textbf{Generalization evaluation.} We use two metrics to evaluate the generalization performance: test set accuracy and accuracy on corrupted data (CIFAR-10-C).

\textbf{Baseline method.} In Sec.\ref{sec:main-results}, we compare DRL with basic AT methods, including PGD-AT \cite{madry2017towards}, TRADES \cite{zhang2019theoretically}, EAT \cite{tramer2017ensemble}, and FAT \cite{wong2020fast}. In Sec.\ref{sec:comparing-with-sota-at}, we show that when combined with other common techniques \cite{hendrycks2019augmix, wang2021augmax}, the DRL can even beat SOTA AT methods \cite{rebuffi2021fixing,wang2023better} in defending transfer-based black-box attacks.

\textbf{Hyper-parameters.} In Sec.\ref{sec:main-results}, we collect 150,000 images as the DRL dataset, consisting of 50,000 original clean images and 100,000 transferable adversarial examples. In Sec.\ref{sec:comparing-with-sota-at}, we collect 200,000 images as the DRL dataset, including 50,000 original clean training images, 50,000 clean images generated from the diffusion model, and their corresponding adversarial images. We train the DRL models for 30 epochs starting from normally trained checkpoints. For each epoch, we select only 50,000 image pairs (clean image and its adversarial image) with the proposed data selection method. The learning rate is 0.1 for WideResNet-28-10 and 0.01 for PreActResNet-18. The optimizer is SGD, with momentum being 0.9 and weight decay being 1e-4. The batch size is 256. We use logarithmic grid search to select the best $\lambda$ in the loss function Eqn. \ref{eqn:naive-DRL}. The searched values of $\lambda$ are shown in Tab.\ref{tab:search-lambda}.

\begin{table}[h]
\centering
\begin{tabular}{ccc}
\toprule
Model                      & AR Loss       & lambda \\
\midrule
\multirow{3}{*}{WRN-28-10} & L1 norm       & 0.001  \\
                           & L2 norm       & 0.0001 \\
                           & KL divergence & 0.001  \\
\midrule
\multirow{3}{*}{Pre-Res18} & L1 norm       & 0.001  \\
                           & L2 norm       & 0.0001 \\
                           & KL divergence & 0.001  \\
\bottomrule
\end{tabular}
\caption{Searched optimal values of $\lambda$ for each model and AR loss on CIFAR-10.}
\label{tab:search-lambda}
\end{table}

\subsection{DRL: Comparing with Various AT Methods}
\label{sec:main-results}

\begin{table*}[]
\centering
\begin{tabular}{cccccccc}
\toprule
\multirow{2}{*}{Model}     & \multirow{2}{*}{Defense} & \multirow{2}{*}{Clean} & \multicolumn{1}{c}{In-box} & \multicolumn{3}{c}{Out-of-box} & \multirow{2}{*}{Data Amount} \\
                           &                          &                        & PGD              & C\&W            & QAA                & Ens Attack        &                              \\
\midrule
\multirow{5}{*}{WRN-28-10} & No                       & \textbf{93.85}         & 49.46            & 55.51           & 11.81              & 10.18             & 0.05M                        \\
                           & PGD-AT                   & 84.66                  & 84.11            & 84.05           & 82.88              & 82.69             & 5M                           \\
                           & TRADES                   & 83.99                  & 83.02            & 83.13           & 82.12              & 81.98             & 5M                           \\
                           & EAT                      & 81.10                  & 80.85            & 81.18           & 82.44              & 82.56             & 5M                           \\
                           & DRL                      & 83.83                  & \textbf{88.60}   & \textbf{90.16}  & \textbf{88.61}     & \textbf{87.06}    & 0.15M                        \\
\midrule
\multirow{6}{*}{Pre-Res18} & No                       & \textbf{91.68}         & 59.44            & 64.21           & 19.50              & 17.58             & 0.05M                        \\
                           & PGD-AT                   & 78.88                  & 78.26            & 78.11           & 77.07              & 77.13             & 5M                           \\
                           & TRADES                   & 79.39                  & 78.45            & 78.56           & 77.37              & 77.44             & 5M                           \\
                           & EAT                      & 78.43                  & 78.05            & 77.83           & 77.70              & 77.21             & 5M                           \\
                           & FAT                      & 83.34                  & 82.54            & 82.43           & 80.92              & 80.95             & 0.15M                        \\
                           & DRL                      & 86.46                  & \textbf{88.35}   & \textbf{89.09}  & \textbf{88.75}     & \textbf{87.35}    & 0.15M    \\
\bottomrule
\end{tabular}
\caption{Comparing the DRL with various adversarial training methods on CIFAR-10. This table reports the clean accuracy (\%) and robust accuracy (\%) of each defense on various transfer attacks, including in-box attack (PGD) and three out-of-box attacks (C\&W, QAA, Ens Attack). The data amount used in training is also listed in this table, which is proportional to the training complexity. The best results are in \textbf{bold}.}
\label{tab:main-results}
\end{table*}

\begin{table*}[t]
\centering
\begin{tabular}{cccccc}
\toprule
\multirow{2}{*}{Model}       & \multirow{2}{*}{Defense} & \multicolumn{2}{c}{Generalization}        & \multicolumn{2}{c}{Robustness}            \\
                             &                          & Clean               & Corrupted           & QAA             & Ens Attack             \\
\midrule
\multirow{6}{*}{WRN-28-10}   & AT-SOTA                  & 92.44/4.99          & 83.49/8.15          & 90.85/6.08          & 90.74/6.11          \\
                             & DA                       & 96.39/2.22          & 89.34/5.05          & 33.18/8.93          & 20.45/9.13          \\
                             & AugMix                   & 96.40/2.30          & 90.35/4.35          & 48.50/8.95          & 31.38/9.81          \\
                             & XDRL                     & 95.48/1.86          & 80.27/7.61          & 93.30/3.67          & 89.00/5.28          \\
                             & XDRL (DA)                  & \textbf{96.74/1.87} & 91.56/4.22          & \textbf{97.13/2.04} & \textbf{95.92/2.58} \\
                             & XDRL (AugMix)              & 96.46/2.13          & \textbf{92.39/4.14} & 96.81/2.31          & 95.64/2.92          \\
\midrule
\multirow{6}{*}{Pre-Res18}   & AT-SOTA                  & 87.35/8.04          & 78.51/10.01         & 85.68/8.88          & 85.77/8.77          \\
                             & DA                       & 94.60/3.00          & 85.09/6.61          & 21.36/7.64          & 16.16/7.06          \\
                             & AugMix                   & 94.83/2.80          & 88.01/5.41          & 37.14/9.82          & 27.07/9.66          \\
                             & XDRL                     & 91.87/4.15          & 78.34/8.43          & 90.64/4.83          & 86.73/6.00          \\
                             & XDRL (DA)                  & 94.77/3.41          & 89.40/5.00          & \textbf{94.64/3.74} & 92.49/4.45          \\
                             & XDRL (AugMix)              & \textbf{95.22/3.09} & \textbf{90.60/4.74} & 94.44/3.55          & \textbf{92.74/4.54} \\
\midrule
\multirow{3}{*}{Res18-DuBIN} & AugMax                   & \textbf{95.77/2.69} & 90.36/4.81          & 63.70/9.68          & 50.62/11.27         \\
                             & XDRL (DA)                  & 95.47/2.76          & 90.85/4.59          & \textbf{96.10/2.53} & 94.55/3.11          \\
                             & XDRL (AugMax)              & 95.70/2.81          & \textbf{91.85/4.49} & 95.68/2.91          & \textbf{94.80/3.13}
                             \\
\bottomrule
\end{tabular}
\caption{Comparing XDRL with AT-SOTA  and data augmentation methods on CIFAR-10. This table reports the robust accuracy (\%) / class-wise standard deviation (\%) of each defense on clean data, corrupted data, and transferable adversarial data (QAA, Ens Attack). The class-wise standard deviation is used to measure robust fairness and lower is fairer. The best results are in \textbf{bold}.}
\label{tab:XDRL-results}
\end{table*}

The experimental results comparing DRL with various AT methods are reported in Tab.\ref{tab:main-results}. We report both clean accuracy and robust accuracy against transfer attacks. One in-box transfer attack (PGD) and three strong out-of-box transfer attacks (C\&W, QAA, and Ens Attack) are adopted to comprehensively verify the black-box robustness of DRL against both seen and unseen attacks. We measure the computational overhead of ATs and DRL with generated data amount. For instance, PGD-AT generates 50,000 images for each epoch and trains for 100 epochs, the generated data amount is 5M (million).
From Tab.\ref{tab:main-results}, we can see that
DRL achieves higher robust accuracy against both in-box and unseen attacks than AT methods. DRL also has higher test set accuracy compared to AT methods. Besides, the less training data amount and faster convergence speed (30 epochs) of DRL lowers the training complexity, making DRL a more affordable solution than AT in realistic applications.

\begin{table}[]
\centering
\begin{tabular}{>{\cellcolor{white}}c|ccc}
\toprule
Model                      & Defense       & Square-100 & Square-2500 \\
\midrule
                           & No            & 48.85      & 0.17        \\
                           & AAA-Linear    & 74.87      & 71.76       \\
                           & DRL            & 58.69      & 1.91        \\
 \multirow{-4}{*}{WRN}& DRL+AAA-Linear & \textbf{80.82}      & \textbf{78.56}       \\
\midrule
                           & No            & 38.42      & 0.09        \\
                           & AAA-Linear    & 62.97      & 56.39       \\
                           & DRL            & 45.39      & 4.22        \\
\multirow{-4}{*}{Res18} & DRL+AAA-Linear & \textbf{77.40}      & \textbf{76.91}       \\
\bottomrule
\end{tabular}
\caption{Combining DRL with defense against query-based attacks on CIFAR-10. The model is WRN (WRN-28-10) and Res18 (PreActRes-18). This table reports the robust accuracy (\%) of the target model under query-based attack. Higher robust accuracy is better. The best results are in \textbf{bold}.}
\label{tab:query-defense}
\end{table}

\subsection{XDRL: Combining DRL with Other Common Techniques}
\label{sec:comparing-with-sota-at}

This section compares XDRL with SOTA AT and data augmentation methods in Tab.\ref{tab:XDRL-results}. For SOTA AT methods, we follow the RobustBench leaderboard \cite{croce2020robustbench} and select top-1 AT methods for WRN-28-10 \cite{wang2023better} and Pre-Res18 \cite{rebuffi2021fixing}, respectively. For combined data augmentation methods, we select DA (naive data augmentation that optimizes the corrupted data with CE loss), AugMix \cite{hendrycks2019augmix}, and AugMax \cite{wang2021augmax}. From the results in Tab.\ref{tab:XDRL-results}, we have the following remarks.

\textbf{Remark 1: DRL outperforms SOTA AT methods when combined with common techniques.} We can see that on each model, XDRL (DA) and XDRL (AugMix) achieve much higher robust accuracy (+5\% $\sim$ 9\%) against strong unseen transfer attacks compared to AT-SOTA. Please note that the computational overhead of XDRL is much less than AT-SOTA, and all those techniques are also commonly adopted in AT-SOTA. These results highlight the advantage of the one-shot training paradigm of DRL in achieving more affordable robustness against realistic transfer-based black-box attacks.

\textbf{Remark 2: DRL further improves the generalization capability and robust fairness when combined with common techniques, such as data augmentations.} For example, when combined with DA, AugMix, and AugMax, XDRL can further enhance the model generalization on unseen corrupted data (+2\% compared to DA and AugMix, +1.5\% compared to AugMax), and the test set generalization capability and the corrupted data generalization capability of XDRL is much higher than that of AT-SOTA. These results can be interesting because adversarial robustness and model generalization are thought to conflict with each other by previous works \cite{wang2021augmax}. Another tricky shortage of AT is its class-wise unfairness \cite{xu2021robust}. The results in Tab.\ref{tab:XDRL-results} show that XDRL has lower class-wise standard deviation, in other words, better robust fairness compared to AT-SOTA.

\subsection{Towards Comprehensive Black-box Defenses}
In this section, we aim to verify the claim that combining DRL with defenses against query-based attacks can lead to comprehensive defenses under black-box settings. To achieve this, we combine DRL with AAA (Adversarial Attack on Attackers) \cite{chen2022adversarial}, a SOTA defense against query-based attacks, and we evaluate its performance on CIFAR-10. We use Square attack \cite{andriushchenko2020square} with 100 and 2500 queries to test the robustness, and the attack perturbation is 8/255. The results are shown in Tab.\ref{tab:query-defense}. We can see that compared to the normally trained model, the DRL model has higher robustness against query-based attacks, although it is not specifically designed to defend against such attacks. Furthermore, when combined with AAA, the defense effectiveness can be significantly enhanced, especially on Pre-Res18, where DRL enhances the robust accuracy by 20.52\% under Square-2500 compared to simply using AAA. These results demonstrate that DRL can be combined with SOTA defenses against query-based attacks to achieve comprehensive black-box defenses.

\subsection{Robustness under Adaptive Attacks}
\label{sec:adaptive-evaluation}

\begin{table*}[]
\centering
\begin{tabular}{c|ccccccc}
\toprule
Model                      & Defense      & Know $M$ & Know $D'$ & Know $\mathcal{L}$ & $M\&D'$ & $M\&\mathcal{L}$  & $D'\& \mathcal{L}$ \\
\midrule
\multirow{2}{*}{WRN-28-10} & DRL     & 90.71  & 53.12   & 91.12  & 62.55 & 92.24 & 34.69 \\
                           & XDRL (DA)  & 97.29  & 76.37   & 95.02  & 75.30 & 94.52 & 31.47 \\
\midrule
\multirow{2}{*}{Pre-Res18} & DRL     & 86.32  & 67.55   & 89.41  & 59.32 & 82.52 & 42.51 \\
                           & XDRL (DA)  & 94.36  & 69.36   & 91.14  & 67.30 & 91.03 & 39.31 \\
\bottomrule
\end{tabular}
\caption{Robust accuracy (\%) of DRL models under six adaptive attacks where the adversary has knowledge on the $M$, $D'$, $\mathcal{L}$, $M\&D'$, $M\& \mathcal{L}$, or $D'\& \mathcal{L}$. The dataset is CIFAR-10 and the attack is PGD with $\epsilon=8/255$.}
\label{tab:adaptive-cifar10}
\end{table*}

As evaluated in Sec.\ref{sec:main-results} and Sec.\ref{sec:comparing-with-sota-at}, DRL has satisfying robustness under the threat model 1 (Realistic Attack). However, it is also valuable to know how DRL performs when the realistic assumption is violated. Specifically, we consider the adversary's knowledge of the clean training dataset $\mathcal{D}$, the augmented training dataset $\mathcal{D}'$, the training objective $\mathcal{L}$, and the model architecture $M$. We consider one realistic attack threat model and six adaptive attack threat models, which are divided in terms of the adversary's knowledge.
\begin{definition}[Adaptive Attacks]
Considering the adversary's knowledge of $M$, $\mathcal{D}'$, and $\mathcal{L}$, stronger adaptive attacks can be divided into six settings where the adversaries know $M$, or $\mathcal{D}'$, or $\mathcal{L}$, or $M \& \mathcal{D}'$, or $M \& \mathcal{L}$, or $\mathcal{D}' \& \mathcal{L}$.
\end{definition}

This section studies the robustness of DRL and XDRL (DA) under six adaptive attacks defined in threat model 2, and the results are reported in Tab.\ref{tab:adaptive-cifar10}. We can see that the knowledge about $D'$ is the most significant factor on the attack success rates. As long as the adversary does not know the proposed DRL dataset, all models evaluated in Tab.\ref{tab:adaptive-cifar10} have a robust accuracy of over 82.52\%. When the adversary has adaptive knowledge on $D'$, the lowest robust accuracy is 53.12\%. These results indicate that DRL still has robustness to some extent under these adaptive black-box attacks. Although DRL only has 31.47\% $\sim$ 42.51\% robust accuracy when the adversary knows both $D' \& \mathcal{L}$, it is an extreme case that is very close to the white-box scenario and may not occur in real-world black-box settings. In conclusion, DRL still has acceptable robustness against adaptive black-box attacks as long as the adversary does not know too much.

\subsection{Ablation Study}
\label{sec:ablation}

\begin{figure*}[t]
    \centering
    \captionsetup[subfigure]{labelformat=empty}
    \begin{minipage}{1\textwidth}
        \begin{subfigure}{.24\textwidth}
            \centering
            \includegraphics[width=1\textwidth]{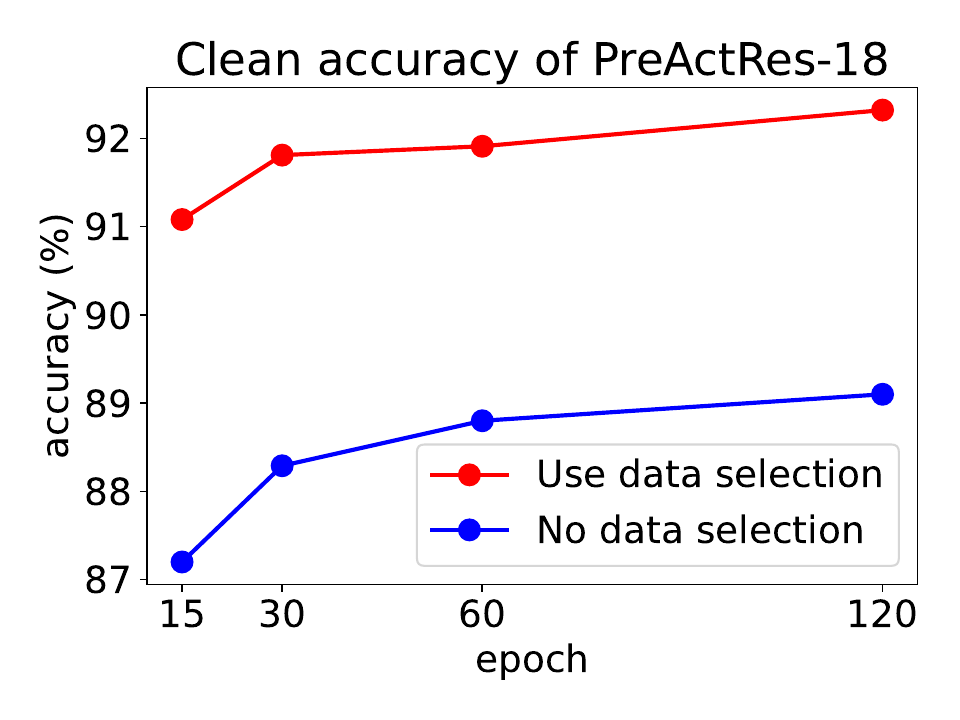}
        \end{subfigure}
        \begin{subfigure}{.24\textwidth}
            \centering
            \includegraphics[width=1\textwidth]{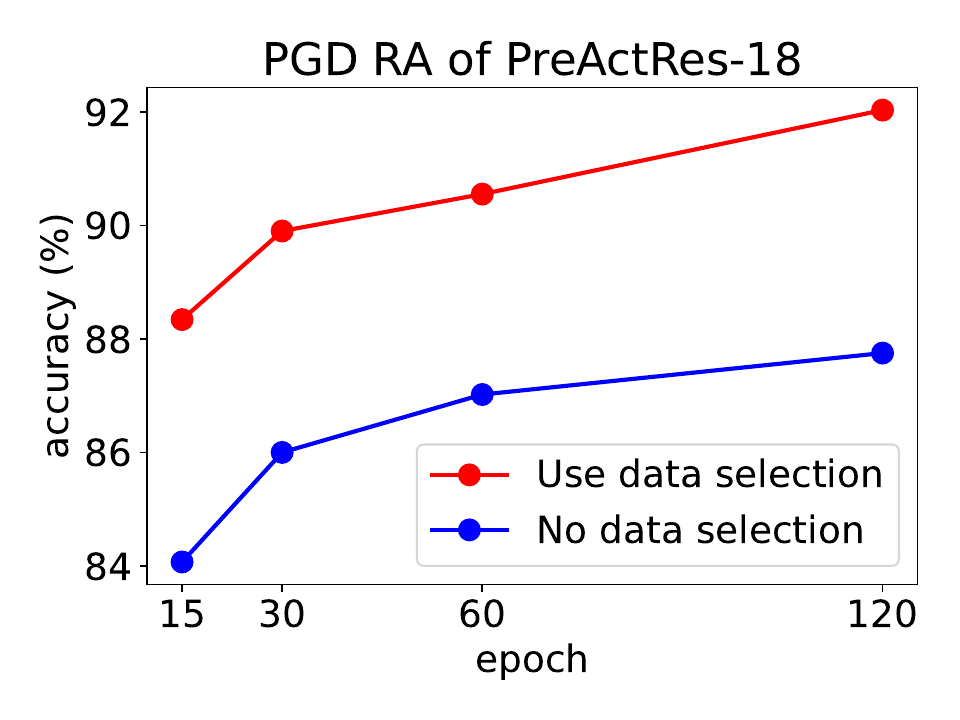}
        \end{subfigure}
        \begin{subfigure}{.24\textwidth}
            \centering
            \includegraphics[width=1\textwidth]{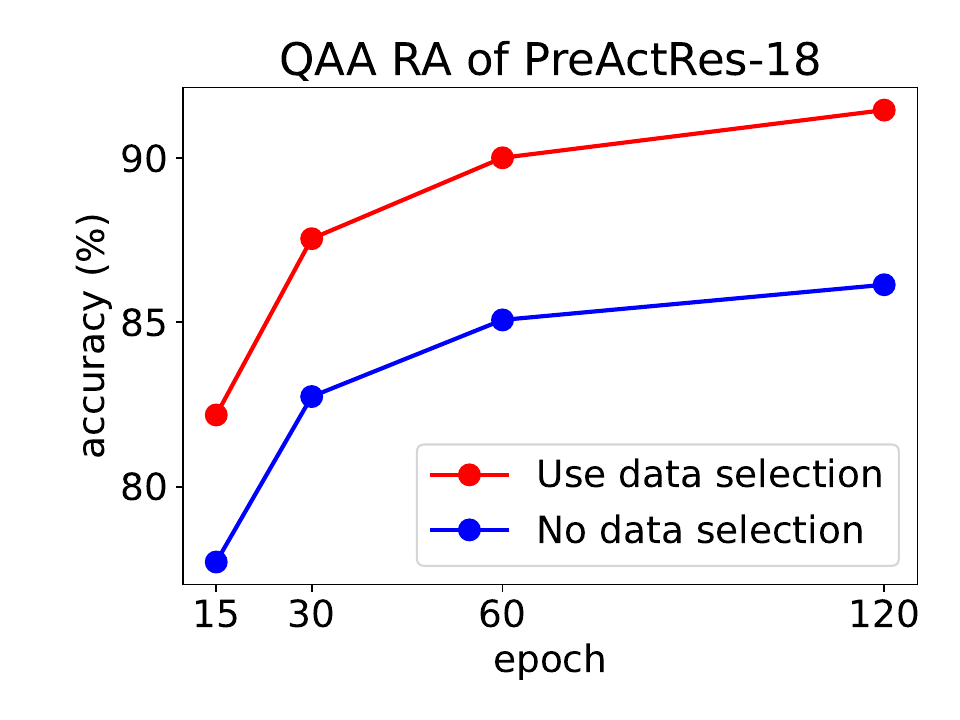}
        \end{subfigure}
        \begin{subfigure}{.24\textwidth}
            \centering
            \includegraphics[width=1\textwidth]{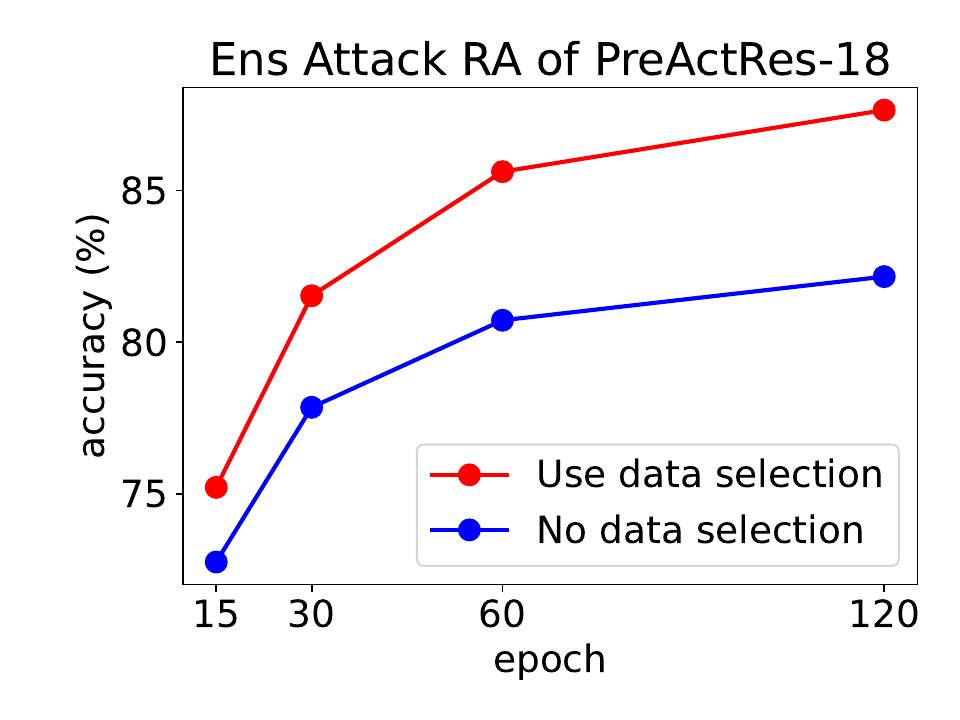}
        \end{subfigure}
    \end{minipage}

    \begin{minipage}{1\textwidth}
        \begin{subfigure}{.24\textwidth}
            \centering
            \includegraphics[width=1\textwidth]{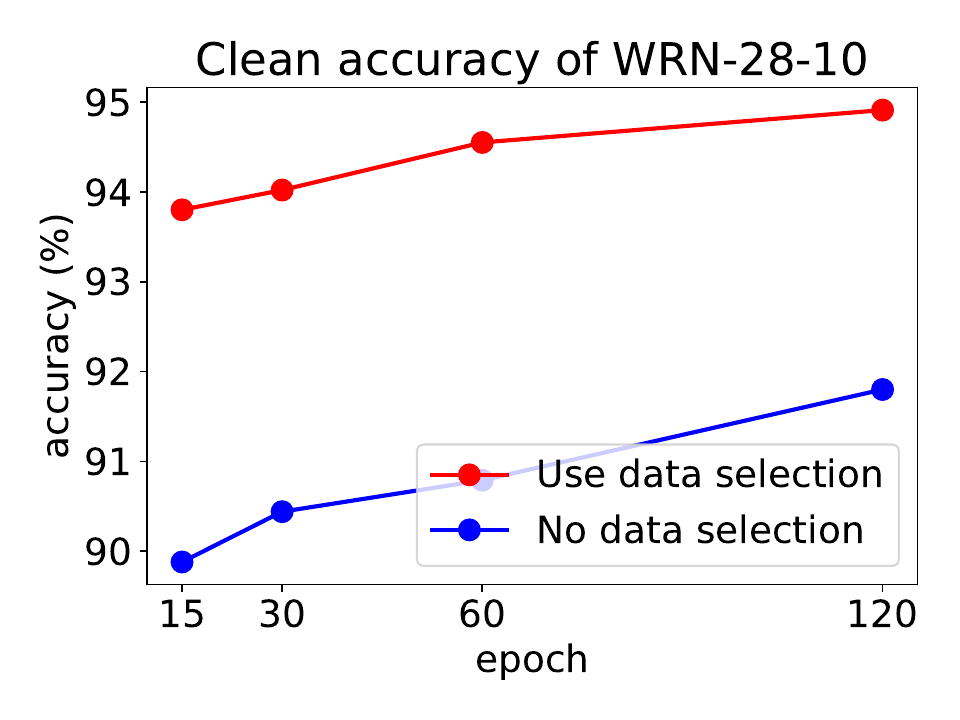}
        \end{subfigure}
        \begin{subfigure}{.24\textwidth}
            \centering
            \includegraphics[width=1\textwidth]{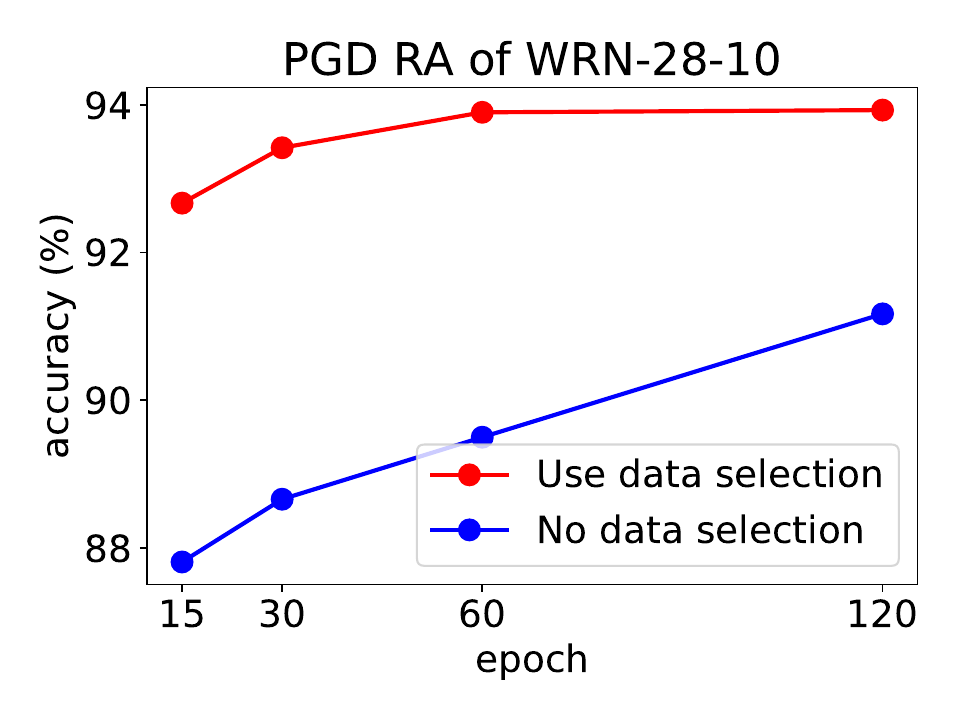}
        \end{subfigure}
        \begin{subfigure}{.24\textwidth}
            \centering
            \includegraphics[width=1\textwidth]{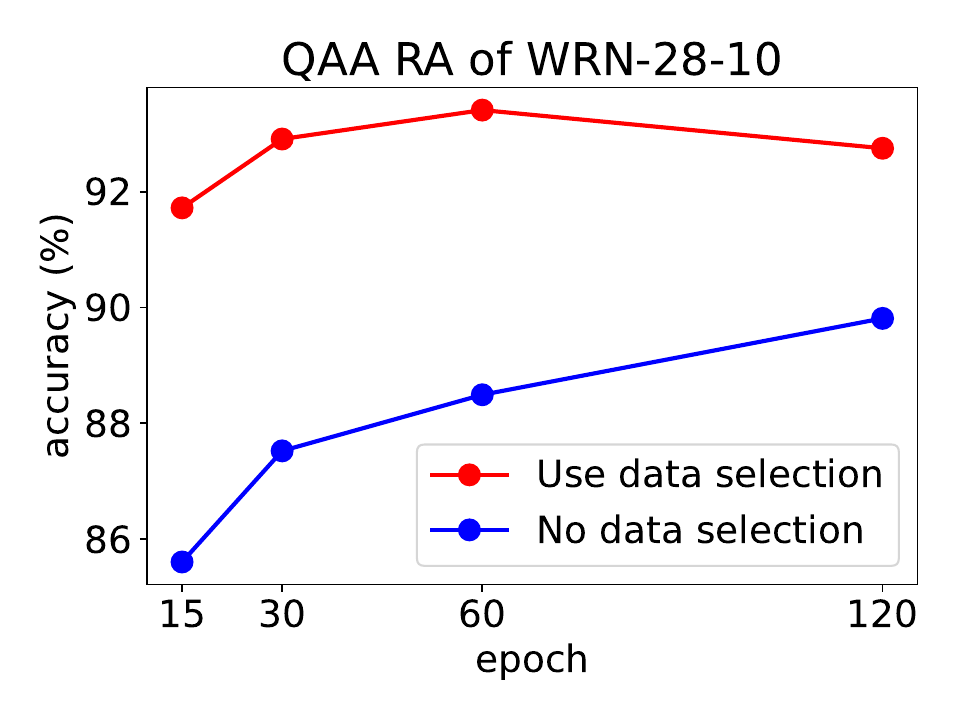}
        \end{subfigure}
        \begin{subfigure}{.24\textwidth}
            \centering
            \includegraphics[width=1\textwidth]{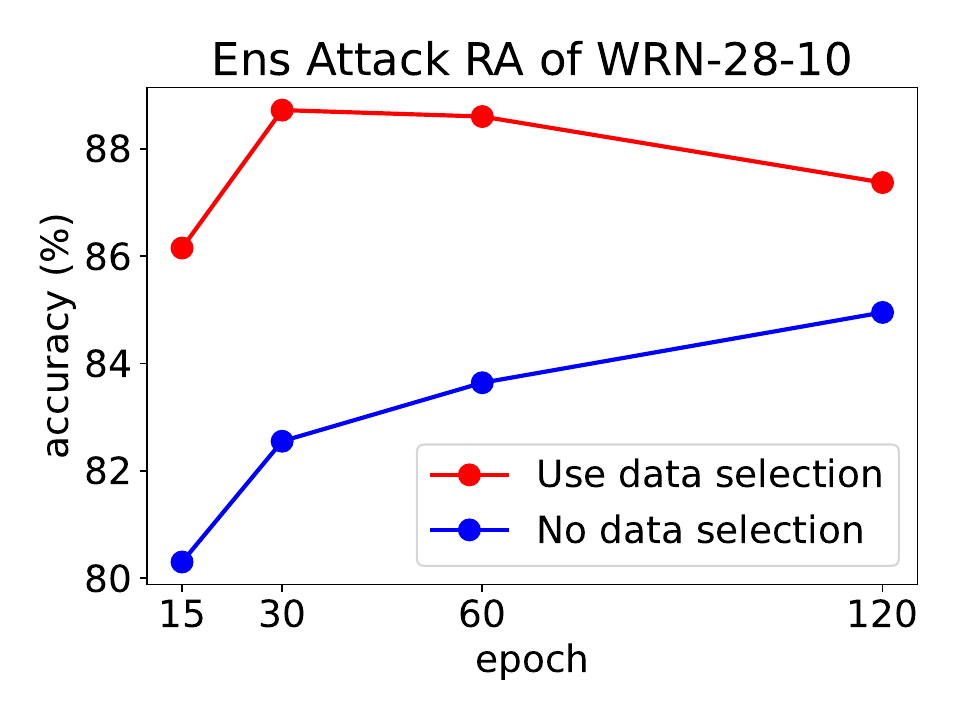}
        \end{subfigure}
    \end{minipage}
    \caption{Ablation results on the data selection. This figure reports the clean accuracy and robust accuracy (RA) of each model using or not using data selection trained with the different number of epochs. The dataset is CIFAR-10. This experiment is repeated three times and the maximum standard deviation is 0.7\%.}
    \label{fig:ablation_data_selection}
\end{figure*}

\textbf{Ablation on the data selection method.} We conduct this ablation experiment to verify whether the proposed data selection method is effective in accelerating the convergence of the DRL model. We train the model with or without data selection methods under different training epochs and compare their clean accuracy and robust accuracy. We set the number of training examples used in each epoch as 50,000 for a fair comparison. The baseline models randomly select training examples in each epoch. The hyper-parameters are consistent with Tab.\ref{tab:main-results} except for the varying training epoch. The results are reported in Fig.\ref{fig:ablation_data_selection}. We can see that all the models using data selection achieve higher accuracy on both clean examples and transferable adversarial examples, compared to their baseline methods not using data selection. Given the same training epoch, the proposed data selection is able to select training examples that the models are not currently ``familiar'' with, thus preventing the models from overfitting.

\textbf{Ablation on the alignment regularization loss function.} This ablation experiment study how to select the alignment regularization (AR) term in Eqn.\ref{eqn:naive-DRL}. We consider three common regularization terms, including the $l_1$-norm, the squared $l_2$-norm, and the KL-divergence. The results are reported in Tab.\ref{tab:ablation_loss}. For each AR loss, we perform the logarithm grid search to select the best $\lambda$. From the results, we can see that the $l_1$-norm achieves the best clean accuracy and robustness. Especially on the strongest transfer attack Ens Attack, the $l_1$-norm outperforms other AR terms by 0.28\% and 1.06\% on WRN-28-10 and Pre-Res18, respectively. Therefore, we select the $l_1$-norm as the AR loss in the DRL framework. Please note that these results differ from the field of data augmentation \cite{wang2022toward}, where the squared $l_{2}$-norm is considered the best.

\begin{table}[h]
\centering
\begin{tabular}{cccccc}
\toprule
Model                      & AR Loss       & Clean & PGD   & QAA   & Ens. \\
\midrule
\multirow{3}{*}{WRN} & $l_{1}$       & \textbf{95.48} & 93.63 & \textbf{93.30} & \textbf{89.00}      \\
                           & squared $l_{2}$       & 94.02 & 93.42 & 92.91 & 88.72      \\
                           & KL div. & 94.06 & \textbf{93.66} & 92.93 & 88.37      \\
\midrule
\multirow{3}{*}{Res18} & $l_{1}$       & \textbf{91.87} & \textbf{91.28} & 90.64 & \textbf{86.73}      \\
                           & squared $l_{2}$       & 91.81 & 89.90 & 87.54 & 81.53      \\
                           & KL div. & 91.81 & 90.87 & \textbf{90.95} & 85.67 \\
\bottomrule
\end{tabular}
\caption{Ablation on the AR loss. This table reports the robust accuracy (\%) of each model using different AR losses against each transfer attack. The dataset is CIFAR-10. The model is WRN-28-10 (WRN) and Pre-Res18 (Res18) This experiment is repeated three times, and the maximum standard deviation is 0.6\%.}
\label{tab:ablation_loss}
\end{table}

\section{Conclusion and Future Work}
This paper finds that transfer-based black-box attacks can be easily mitigated by the one-shot training paradigm, which can be a more affordable solution than adversarial training in realistic settings. Based on this insight, this paper proposes the Data-centric Robust Learning (DRL) method consisting of adversarial data augmentation and data selection mechanisms to defend against transfer-based black-box attacks. Extensive experimental results show that compared to adversarial training, DRL is more efficient, effective, and fair in defending transfer-based black-box attacks. DRL can be combined with common data augmentation techniques and outperform the SOTA AT method in defending transfer-based black-box adversarial attacks. When combined with other black-box defenses, DRL achieves comprehensive black-box defense. Despite the contribution of this paper, how to generate high-quality datasets to enhance the robustness of DNNs is still an open question for future work.

\bibliography{aaai24}

\end{document}